\documentclass[11pt]{article}
\usepackage[english]{babel}
\usepackage{amssymb,amsmath}

\newtheorem{definition}{Definition}
\newtheorem{proposition}{Proposition}
\newtheorem{theorem}{Theorem}
\newtheorem{remark}{Remark}

\newcommand{\tr}{\mathrm{tr}}

\newcommand{\spa}{\hspace{-2mm}}
\newcommand{\tre}{\hspace{0.3mm}}
\newcommand{\quattro}{\hspace{0.4mm}}
\newcommand{\cinque}{\hspace{0.5mm}}
\newcommand{\sei}{\hspace{0.6mm}}
\newcommand{\sette}{\hspace{0.7mm}}
\newcommand{\otto}{\hspace{0.8mm}}

\newcommand{\mtre}{\hspace{-0.3mm}}

\newcommand{\errep}{\mathbb{R}^{\mbox{\tiny $+$}}}

\newcommand{\erre}{\mathbb{R}}

\newcommand{\ccc}{\mathbb{C}}
\newcommand{\conv}{\star}

\newcommand{\de}{\mathrm{d}}

\newcommand{\bsp}{\mathcal{J}}
\newcommand{\bspd}{\mathcal{J}^{\ast}}
\newcommand{\bspdd}{\mathcal{J}^{\ast\ast}}
\newcommand{\csg}{\mathfrak{C}}
\newcommand{\opa}{\hat{A}}
\newcommand{\opb}{\hat{B}}
\newcommand{\ops}{\hat{S}}
\newcommand{\opsp}{\hat{S}^\prime}

\newcommand{\cz}{\mathsf{C}_0}

\newcommand{\dom}{\mathrm{Dom}}

\newcommand{\hh}{\mathcal{H}}

\newcommand{\smis}{\mathcal{M}^1(G)}

\newcommand{\twis}{\mathfrak{S}_{t}}
\newcommand{\twishs}{\check{\mathfrak{S}}_{t}}

\newcommand{\dwis}{\mathfrak{D}_{t}}

\newcommand{\bH}{\mathcal{B}(\mathcal{H})}

\newcommand{\mut}{\mu_t}
\newcommand{\mus}{\mu_s}

\newcommand{\ran}{\mathrm{Ran}\hspace{0.3mm}}

\newcommand{\sunig}{\mathrm{SU}\hspace{0.3mm}}

\newcommand{\prop}{\mathfrak{P}}

\newcommand{\sopa}{\mathfrak{A}}

\newcommand{\trc}{\mathcal{B}_1(\mathcal{H})}

\newcommand{\intG}{\int_G}
\newcommand{\mG}{{\nu_G}}
\newcommand{\ldg}{\mathrm{L}^2(G)}
\newcommand{\lgc}{\mathrm{L}^2(G,\mG;\mathbb{C})}

\newcommand{\emme}{\mathsf{m}\hspace{0.3mm}}
\newcommand{\temme}{\breve{\mathsf{m}}\hspace{0.3mm}}
\newcommand{\twoside}{\mathcal{T}_{\emme}\hspace{-0.2mm}}

\newcommand{\wlim}{\mbox{w-}\hspace{-0.7mm}\lim}

\newcommand{\slim}{\mbox{s\hspace{0.3mm}-}\hspace{-0.7mm}\lim}

\newcommand{\hs}{\mathcal{B}_2(\mathcal{H})}

\newcommand{\wigu}{\mathcal{W}}

\newcommand{\smg}{\mathfrak{T}_t^{\hspace{0.2mm}\emme}}

\newcommand{\urep}{{U\hspace{-0.5mm}\vee\hspace{-0.5mm} U}}
\newcommand{\rep}{{\underline{U\hspace{-0.5mm}\vee\hspace{-0.5mm} U\hspace{-0.8mm}}\hspace{0.8mm}}}

\newcommand{\supops}{\mathcal{L}(\mathcal{H})}
\newcommand{\dsupops}{\mathcal{L}^\prime(\mathcal{H})}

\newcommand{\modu}{\Delta_G}
\newcommand{\uni}{\mathfrak{V}}
\newcommand{\unim}{{\mu[\uni]}}
\newcommand{\unime}{{\mu[\uni]^{\mathrm{e}}}}
\newcommand{\unimd}{{\mu[\uni]}^{\ast}}
\newcommand{\unimt}{{\mut[\uni]}}
\newcommand{\unimo}{{\mu_0[\uni]}}
\newcommand{\unims}{{\mus[\uni]}}
\newcommand{\unimts}{{\mu_{t+s}[\uni]}}
\newcommand{\unimtd}{{\mut[\uni]}^{\ast}}
\newcommand{\umt}{{\mut[U]}}
\newcommand{\umtd}{{\mut[U]}^{\ast}}
\newcommand{\urepmt}{{\mut[\urep]}}
\newcommand{\urepmtd}{{\mut[\urep]}^{\ast}}
\newcommand{\repmt}{{\mut[\rep]}}
\newcommand{\twomt}{{\mut[\twoside]}}

\newcommand{\effe}{\phi}
\newcommand{\deffe}{F}
\newcommand{\elleu}{\mathrm{L}^1}
\newcommand{\petint}{\diamond\hspace{-1.5mm}\int}
\newcommand{\bsrep}{T}
\newcommand{\bbsp}{\mathcal{B}(\bsp)}
\newcommand{\bbspd}{\mathcal{B}(\bspd)}
\newcommand{\plangle}{\left (}
\newcommand{\prangle}{\right )}
\newcommand{\co}{\mathrm{co}\tre}
\newcommand{\clco}{\overline{\mathrm{co}}\tre}
\newcommand{\mset}{\mathcal{E}}
\newcommand{\nulset}{\mathcal{E}_{\effe}}

\begin{document}

\title{On a certain class of semigroups of operators}

\author{
Paolo Aniello$^{\hspace{0.6mm}1,2,3,4}$ \vspace{2mm}\\
\small \it $^1$ Dipartimento di Scienze Fisiche dell'Universit\`a
di Napoli `Federico II', \\
\small \it Complesso Universitario di Monte S.\ Angelo, via Cintia,
I-80126 Napoli, Italy
\\ \small \it $^2$ Istituto Nazionale di Fisica Nucleare (INFN), Sezione di Napoli, Napoli, Italy  \\
\small \it $^3$ MECENAS, Universit\`a
di Napoli `Federico II', Napoli, Italy \\
\small \it $^4$ Facolt\`a di Scienze Biotecnologiche,
Universit\`a di Napoli `Federico II', Napoli, Italy
}

\maketitle

\begin{abstract}
We define an interesting class of semigroups of operators in Banach spaces,
namely, the \emph{randomly generated semigroups}. This class contains as a remarkable
subclass a special type of quantum dynamical semigroups introduced by Kossakowski in the
early 1970s. Each randomly generated semigroup is associated, in a natural way, with a pair
formed by a representation or an antirepresentation of a locally compact group
in a Banach space and by a convolution
semigroup of probability measures on this group. Examples of randomly generated semigroups
having important applications in physics are briefly illustrated.
\end{abstract}

\section{Introduction}
\label{intro}

In the early 1970s, Kossakowski~{\cite{Kossakowski}} introduced an interesting class
of semigroups of operators, or more precisely --- according to the
standard terminology in use nowadays~{\cite{Holevo}} ---
of \emph{quantum dynamical semigroups}. In particular, the work of Kossakowski established
a remarkable link between the theory of Brownian motion~{\cite{Nelson,Ito,Yosida,Hunt}}
and the theory of open quantum systems~{\cite{Breuer}.

In a recent paper~{\cite{Aniello-brownian}}, the mentioned class of quantum dynamical semigroups
--- the so-called \emph{twirling semigroups} ---
has been studied in detail. In particular, in the case of finite-dimensional
open quantum systems a complete characterization
of the infinitesimal generators of the twirling semigroups
associated with representations of Lie groups has been obtained.
In the infinite-dimensional case, thanks to Nelson's theory
of analytic vectors~{\cite{Nelson-anv}}, one can extend some of the
results of~{\cite{Aniello-brownian}} by taking
care of the domains of the (in general, unbounded) infinitesimal
generators of the twirling semigroups. This task will be
accomplished elsewhere~{\cite{Aniello-progress}}.

The twirling semigroups arise in the study of various
physical contexts. For instance, the analysis of the infinitesimal generators of
the twirling semigroups shows that this class of semigroups of (super)operators
includes
the semigroups describing the dynamics of a
finite-dimensional system with a purely random Gaussian stochastic
Hamiltonian~{\cite{Go-Kos}}, and the reduced dynamics of a
finite-dimensional system in the limit of singular coupling to a
reservoir at infinite temperature~{\cite{Fri-Go}}. Moreover, the twirling semigroups
turn out to be relevant in connection with applications to quantum information theory; see~{\cite{Aniello-brownian}}
and references therein.

The main aim of the present contribution is to provide a natural generalization of the class of
semigroups introduced by Kossakowski in his seminal paper~{\cite{Kossakowski}} by defining
the larger class of \emph{randomly generated semigroups}. These semigroups of operators --- as well as
their adjoint semigroups --- are associated, in a straightforward way, with pairs
formed by a representation, or an antirepresentation, of a locally compact group
in a Banach space and by a convolution
semigroup of probability measures on this group.
Although conceptually very simple, this construction involves some technical aspects.
In particular, as it will be clear in the following,
the rigorous definition of randomly generated
semigroups relies on the theory of integration of vector-valued functions.

In addition to its intrinsic interest, we think that the definition of
this larger class of semigroups of operators
is useful in order to better clarify the relevant mathematical structures involved
in the construction of twirling semigroups, also in view of the mentioned extension~{\cite{Aniello-progress}} of
results previously obtained in the recent paper~{\cite{Aniello-brownian}}.
Furthermore, as we will show later on, the notion of randomly generated semigroup
allows us to encompass in a unified framework the twirling semigroups and a related class of
semigroups of operators, namely, the \emph{tomographic semigroups}.

The paper is organized as follows. In~Sect.~{\ref{basic}}, we briefly review
the main mathematical tools involved in our analysis.  Next,
in Sect.~{\ref{random}}, we introduce the notions of `randomly generated operator'
and of `randomly generated semigroup' (of operators).
Some remarkable examples of randomly generated semigroups are discussed in
Sect.~{\ref{examples}}. Finally, in Sect.~{\ref{conclusions}}, a few conclusions are drawn.

\section{Basic facts}
\label{basic}

In this section, we will fix the main notations and recall some
basic technical facts. For further details, the reader may consult the standard references {\cite{Folland}}
(functional analysis and basics in probability theory),
\cite{Hille,Yosida-book} (semigroups of operators), \cite{Diestel} (integration of vector-valued functions),
\cite{Raja} (representation theory) and
\cite{Grenander,Heyer} (probability theory on groups).
We implicitly refer to these sources wherever some notion or result is
mentioned or used without any explicit reference.

Throughout the paper we will consider Banach spaces over the field of real or complex numbers,
usually with no specification of which of the two fields is involved.
Let $\bsp$ be a Banach space. Denoting by
$\errep$ the set of non-negative real numbers, a family
$\{\csg_t\}_{t\in\errep}$ of bounded linear operators in $\bsp$ is
said to be a \emph{continuous semigroup of operators} if the
following conditions hold:
\begin{enumerate}
\item $\csg_t\cinque \csg_s = \csg_{t+s}$, $t,s\ge0$ \ (semigroup property);

\item $\csg_0 = I$;

\item $\lim_{t\downarrow 0}\|\csg_t\cinque\phi -\phi\|=0$,
$\forall\quattro\phi\in\bsp$, i.e., $\slim_{t\downarrow
0}\csg_t=I$ \  (strong right continuity at $t=0$).
\end{enumerate}
Here and in the following, $I$ is the identity operator.
According to a classical result~{\cite{Hille}},
the previous conditions imply that the map $\errep\ni t\mapsto
\csg_t\in\bsp$ is strongly continuous.
The last condition is equivalent to
the assumption that $\wlim_{t\downarrow 0}\csg_t=I$ (weak limit), see~{\cite{Yosida-book}}.
Moreover, it is a well known fact that
a semigroup of operators $\{\csg_t\}_{t\in\errep}$ admits a densely
defined \emph{infinitesimal generator}, namely, the closed linear
operator $\sopa$ in $\bsp$ defined by
\begin{equation}
\dom(\sopa):=\Big\{\phi\in\bsp\colon \exists\cinque
\lim_{t\downarrow 0}
t^{-1}\big(\csg_t\cinque\phi-\phi\big)\Big\},\ \ \
\sopa\cinque\phi := \lim_{t\downarrow 0}
t^{-1}\big(\csg_t\cinque\phi-\phi\big),\
\forall\quattro\phi\in\dom(\sopa).
\end{equation}

Let $(X,\mathcal{S},\mu)$ be a measure space, with $\mu$ a probability measure.
Suppose that
\begin{equation}
X\ni x\mapsto\effe(x)\in\bsp,\ \ \
X\ni x\mapsto\deffe(x)\in\bspd
\end{equation}
--- where, with standard notation, $\bspd$ is the dual space of $\bsp$ ---
are, respectively, a weakly-$\mu$-measurable and weakly$^\ast$-$\mu$-measurable
function such that
\begin{equation}
\plangle E, \effe (\cdot)\prangle,  \plangle\deffe (\cdot) ,\eta\prangle \in\elleu(\mu),\ \ \
\forall\quattro\eta\in\bsp, \forall\quattro E \in\bspd;
\end{equation}
here, $\plangle \cdot ,\cdot \prangle$ is the `pairing' between $\bsp$ and $\bspd$
(the same notation will be adopted for the pairing between $\bspd$ and the double dual $\bspdd$ of $\bsp$).
Then, for any $\mset\in\mathcal{S}$ (in particular, for $\mset = X$), one can define two vectors
\begin{equation} \label{twovecs}
\int_{\mset} \effe(x)\; \de\mu(x)\in\bspdd,\ \ \ \int_{\mset} \deffe(x)\; \de\mu(x)\in\bspd,
\end{equation}
where the integrals in~{(\ref{twovecs})} are a \emph{Dunford integral} and a \emph{Gelfand
integral}~{\cite{Diestel}}, respectively. In the case where the the first integral in~{(\ref{twovecs})}
defines a vector belonging to $\bsp$ (consider the natural injection of $\bsp$ into $\bspdd$) ---
i.e., if the function $X\ni x\mapsto\effe(x)\in\bsp$ is \emph{Pettis integrable relatively to} $\mset$ (this terminology
is nonstandard) --- we will
denote this vector by the symbol
\begin{equation} \label{pett}
\petint_{\mset} \effe(x)\; \de\mu(x)\in\bsp.
\end{equation}
If the function $X\ni x\mapsto\effe(x)\in\bsp$ is Pettis integrable relatively to every $\mset\in\mathcal{S}$, then it is said to be a
\emph{Pettis integrable} function, and the vector~{(\ref{pett})} is the standard \emph{Pettis integral} of this function over $\mset$~{\cite{Diestel}}.
Later on, we will use the fact that under certain conditions the Pettis integral can be replaced
by the ordinary Bochner integral~{\cite{Yosida-book,Diestel}}.

Let $G$ be a locally compact, second countable, Hausdorff
topological group (in short, l.c.s.c.\ group). The symbol $e$ will
denote the identity in $G$. We will say that a map $\bsrep\colon G\rightarrow\bbsp$
--- from the l.c.s.c.\ group $G$ into the Banach algebra $\bbsp$ of bounded linear operators
in the Banach space $\bsp$ --- is a \emph{uniformly bounded representation} of $G$ if it is a weakly
continuous map such that $\sup_{g\in G} \|T(g)\|<\infty$, $\bsrep(e)=I$ and $\bsrep(gh)=\bsrep(g)\,\bsrep(h)$, for all
$g,h\in G$. In the case where, instead, the last relation is replaced by $\bsrep(gh)=\bsrep(h)\,\bsrep(g)$,
we will call the map $\bsrep$ a  \emph{uniformly bounded antirepresentation}. Moreover,
we will mean by the term {\it projective representation} of $G$, in a separable complex
Hilbert space $\mathcal{H}$, a map $U$ of $G$ into
$\mathcal{U}(\mathcal{H})$  --- the unitary group of $\hh$ --- such
that
\begin{itemize}
\item
$U$ is a weakly Borel map, i.e.\ $G\ni g\mapsto
\langle\phi,U(g)\,\psi\rangle\in\mathbb{C}$ is a Borel function, for
any pair of vectors $\phi,\psi\in\mathcal{H}$ (with $\langle \cdot ,\cdot \rangle$ denoting the scalar product);
\item
$U(e)=I$;
\item
denoting by $\mathbb{T}$ the circle group, namely the group of
complex numbers of modulus one, there exists a Borel function $\emme
\colon G\times G\rightarrow\mathbb{T}$ such that
\begin{equation}
U(gh)=\emme (g,h)\,U(g)\,U(h),\ \ \ \forall\hspace{0.5mm} g,h\in G .
\end{equation}
\end{itemize}
The function $\emme$ is called the {\it multiplier associated with}
$U$. Clearly, in the case where $\emme\equiv 1$, $U$ is a
standard unitary representation; in this case, according to a well
known result, the hypothesis that the map $U$ is weakly Borel
implies that it is, actually, strongly continuous.

We will denote by $\smis$ the semigroup --- with respect to
convolution of measures
--- of all Borel probability measures on $G$, \emph{endowed with
the weak topology} (which, in $\smis$, coincides with the vague
topology). For any pair $\mu,\nu\in\smis$ we will denote by $\mu\conv\nu$
the convolution of $\mu$ with $\nu$. The symbol $\delta\equiv\delta_e$ will denote the Dirac
measure at $e$, measure that is, of course, the identity in the
semigroup $\smis$. By a \emph{continuous convolution semigroup of
measures} on $G$ we mean a subset $\{\mut\}_{t\in\errep}$ of $\smis$
such that the map $\errep\ni t\mapsto\mut\in\smis$ is a homomorphism
of semigroups and
\begin{equation}
\lim_{t\downarrow 0} \mut = \delta.
\end{equation}
It is a well known fact that this condition implies that the
homomorphism $t\mapsto\mut$ is continuous.

\section{Randomly generated operators and semigroups}
\label{random}

We will start this section by introducing the notion of \emph{randomly generated operator}.
Let $(X,\mathcal{S},\mu)$ be, as above, a probability space, $\bsp$ a Banach space and
$\uni\colon X\rightarrow \bbsp$ a
(norm) bounded map. Suppose, moreover, that this map is weakly-$\mu$-measurable; namely,
that, for any $\effe\in\bsp$ and $\deffe\in\bspd$, the function
\begin{equation}
X\ni x \mapsto \plangle \deffe, \effe (x)\prangle \in\ccc, \ \ \ \effe (x)\equiv \uni(x)\cinque \effe,
\end{equation}
is $\mu$-measurable. Clearly, this function coincides with
$X \ni x \mapsto \plangle \deffe(x), \effe \prangle\in\ccc$,
$\deffe (x)\equiv \uni(x)^\ast \deffe$, where $\uni(x)^\ast\in\bbspd$ is the adjoint of the operator $\uni(x)$.
Furthermore, since the map $\uni$ is bounded, the function
$x \mapsto \plangle \deffe, \effe (x)\prangle = \plangle \deffe(x), \effe \prangle$
belongs to $\elleu(\mu)$.

We can now actually define three linear operators.
\begin{definition}
The \emph{extended randomly generated operator} associated with the pair $(\uni,\mu)$ is the
linear operator $\unime\colon\bsp\rightarrow\bspdd$ determined by
\begin{equation} \label{extrandom}
\unime \tre \effe = \int_X \effe (x)\; \de\mu(x), \ \ \ \effe (x)\equiv \uni(x)\cinque \effe,
\ \ \ \forall\quattro\effe\in\bsp,
\end{equation}
where the integral in~{(\ref{extrandom})} is a Dunford integral.
The \emph{dual randomly generated operator} associated with the pair $(\uni,\mu)$ is the
linear operator $\unimd\colon\bspd\rightarrow\bspd$ determined by
\begin{equation} \label{dualrandom}
\unimd  \deffe = \int_X \deffe (x)\; \de\mu(x), \ \ \ \deffe (x)\equiv \uni(x)^\ast \deffe,
\ \ \ \forall\quattro\deffe\in\bspd,
\end{equation}
where the integral in~{(\ref{dualrandom})} is a Gelfand integral. Finally, suppose that
--- in addition to the previous assumptions --- for every $\effe\in\bsp$,
the map $X\ni x \mapsto \uni(x)\cinque \effe\in\bsp$ is Pettis integrable relatively to $X$. Then,
the \emph{randomly generated operator} associated with the pair $(\uni,\mu)$ is the
linear operator $\unim\colon\bsp\rightarrow\bsp$ determined by
\begin{equation} \label{randomly}
\unim \cinque \effe  =  \petint_X \effe (x)\; \de\mu(x), \ \ \ \effe (x)\equiv \uni(x)\cinque \effe,
\ \ \ \forall\quattro\effe\in\bsp.
\end{equation}
\end{definition}
Obviously, in the case where the Banach space $\bsp$ is reflexive, the randomly generated operator $\unim$
and the extended randomly generated operator $\unime$ coincide. Also observe that the notation adopted for the
dual randomly generated operator is coherent with the fact that the operator $\unimd$ is the adjoint of $\unim$.

\begin{proposition}
With the previous assumptions and definitions, the randomly, dual randomly and extended randomly generated operators associated
with the pair $(\uni,\mu)$ are bounded.
\end{proposition}

\noindent {\bf Proof:} Let us prove the statement for the extended randomly generated operator $\unime$. The other two cases
are analogous (in the case of the operator $\unim$ use the fact that the natural injection of $\bsp$ into $\bspdd$ is isometric).
In fact, we have that
\begin{eqnarray}
\|\unime\| \spa & = & \spa \sup_{\effe\in\bsp,\sei \|\effe\|=1} \|\unime \tre \effe\|
\nonumber \\
& = & \spa \sup_{\effe\in\bsp,\sei \|\effe\|=1} \otto
\sup_{\deffe\in\bspd \mtre,\sei \|\deffe\|=1}
|\plangle \unime \tre \effe , \deffe \prangle|
\nonumber \\
& \le & \spa
\sup_{\effe\in\bsp,\sei \|\effe\|=1} \otto
\sup_{\deffe\in\bspd \mtre,\sei \|\deffe\|=1} \int_X | \plangle\deffe,\uni (x) \tre \effe\prangle| \;\de\mu(x)
\le \sup_{x\in X} \|\uni(x)\| ,
\end{eqnarray}
where, in the second line, $\plangle \cdot ,\cdot \prangle$ is the pairing between $\bspd$ and $\bspdd$.
The proof is complete.~{$\square$}

We will now show that the additional condition that allows us to define the operator $\unim$ is automatically satisfied under a certain
hypothesis.

\begin{proposition} \label{separably}
Suppose, as above, that $\uni\colon X\rightarrow \bbsp$ is a bounded, weakly-$\mu$-measurable map.
Suppose, moreover, that for every $\effe\in\bsp$ the map $x\mapsto \uni(x)\cinque \effe$
is $\mu$-essentially separably valued, i.e., that there exists $\nulset\in\mathcal{S}$, with
$\mu(\nulset)=0$, such that $\uni(X\smallsetminus\nulset)\cinque\phi$ is a (norm) separable subset of $\bsp$. Then, the randomly generated
operator associated with the pair $(\uni,\mu)$ exists and, for every $\effe\in\bsp$,
\begin{equation} \label{hull}
\unim\cinque\effe\in  \clco (\uni(X)\cinque\effe)\subset\bsp,
\end{equation}
namely, the vector $\unim\cinque\effe$
belongs to the closed convex hull of the set $\uni(X)\cinque\effe\equiv\{\uni(x)\cinque\effe\}_{x\in X}$.
\end{proposition}

\noindent {\bf Proof:} Indeed, under the mentioned hypotheses, by Pettis measurability theorem~{\cite{Diestel}}
the map $x\mapsto \uni(x)\cinque\effe$ is $\mu$-measurable, for any $\effe\in\bsp$.
Furthermore, the function $x\mapsto \|\uni(x)\cinque\effe\|$ belongs to $\elleu(\mu)$.
Therefore, for every $\effe\in\bsp$, the map $x\mapsto \uni(x)\cinque\effe$ is actually Bochner integrable~{\cite{Diestel}}. Then, relation~{(\ref{hull})}
follows as a well known property of the Bochner integral.~{$\square$}

\begin{remark} {\rm
In the case where $\bsp$ is finite-dimensional, clearly we have a stronger result. Let
$\uni\colon X\rightarrow \bbsp$ a weakly-$\mu$-measurable map. Then, $\unim = \int_X \uni (x)\; \de\mu(x)$
(integral of matrix-valued functions), and
\begin{equation} \label{hullbis}
\unim\in  \clco (\uni(X))\subset\bbsp;
\end{equation}
compare with~{(\ref{hull})}.~{$\blacksquare$}
}
\end{remark}

In the case where $\bsp$ coincides with a complex Hilbert space $\hh$ and the range of the map $\uni$ is contained in the unitary group $\mathcal{U}(\hh)$ of
$\hh$, we will call the operator $\unim$ a \emph{random unitary operator}.
\begin{proposition}
If $\hh$ is a finite-dimensional complex Hilbert space, then every random unitary operator $\mathfrak{R}$ in
$\hh$ is of the form
\begin{equation}
\mathfrak{R} = \sum_{k\in\mathcal{K}} p_k\sei \uni_{k},
\end{equation}
where $\mathcal{K}$ is a finite index set, $\{p_k\}_{k\in\mathcal{K}}$ a probability distribution
and $\{\uni_{k}\}_{k\in\mathcal{K}}\subset\mathcal{U}(\hh)$.
\end{proposition}

\noindent {\bf Proof:} The proof is straightforward and it is left to the reader (hint: exploit relation~{(\ref{hullbis})},
the fact that the unitary group $\mathcal{U}(\hh)$ is compact
--- hence, $\co(\mathcal{U}(\hh))=\clco(\mathcal{U}(\hh))\supset\clco (\uni(X))$
 --- and Caratheodory's theorem~{\cite{Stoer}}).~{$\square$}

At this point, we will focus on the case where $X$ coincides with a l.c.s.c.\ group $G$,
endowed with a continuous convolution semigroup of
measures $\{\mut\}_{t\in\errep}$. We will further assume that $\uni$ is a uniformly
bounded representation or antirepresentation of $G$ in a Banach space $\bsp$.
We can now consider two sets of operators, namely, the set $\{\unimtd\}_{t\in\errep}$
and --- assuming that, for every $\effe\in\bsp$,
the map $x \mapsto \uni(x)\cinque \effe$ is Pettis integrable, relatively to $X$, with respect to $\mut$ for all $t>0$  ---
the set $\{\unimt\}_{t\in\errep}$, as well. By Proposition~{\ref{separably}}, the last assumption
is superfluous in the case where the Banach space $\bsp$ is separable.

\begin{theorem} \label{semi-g}
With the previous notations and assumptions, the sets of operators $\{\unimt\}_{t\in\errep}$
and $\{\unimtd\}_{t\in\errep}$ are continuous semigroups of operators in $\bsp$
and $\bspd$, respectively.
\end{theorem}

\noindent {\bf Proof:} We will prove the statement for the set of operators $\{\unimt\}_{t\in\errep}$ only,
since the other case is analogous.

Let us first prove that the set of operators $\{\unimt\}_{t\in\errep}$ enjoys the semigroup property. In order to fix ideas,
assume that $\uni$ is a representation (rather than an antirepresentation).
Then, for any $\effe\in\bsp$ and $\deffe\in\bspd$, we have that
\begin{eqnarray}
\plangle F,\unimt\sette\unims\cinque\effe \prangle \spa & = & \spa \int_G  \de \mut(g)\int_G\de \mus (h)\;
\plangle F,\uni (g h)\cinque\effe \prangle \nonumber \\
& = & \spa \int_G \de \mut\conv\mus (g)\; \plangle F,\uni (g)\cinque\effe \prangle \nonumber \\
& = & \spa
\int_G \de \mu_{t+s} (g)\; \plangle F,\uni (g)\cinque\effe \prangle = \plangle\deffe , \unimts\cinque \effe\prangle .
\end{eqnarray}
Therefore, since the elements of $\bspd$ separate points in $\bsp$,
$\unimt\sette\unims=\unimts$, for all $t,s\in\errep$. It is clear that
the proof in the case where $\uni$ is an antirepresentation runs along the same lines. It is obvious, moreover,
that $\unimo=I$.

Let us now show that the semigroup of operators $\{\unimt\}_{t\in\errep}$ is continuous.
Actually, as recalled in
Sect.~{\ref{basic}}, it suffices to prove the weak right continuity
at $t=0$.
To this aim observe that, for any $\effe\in\bsp$ and $\deffe\in\bspd$, the function
\begin{equation}
G\ni g\mapsto|\plangle\deffe , \uni (g)\cinque\effe\prangle-\plangle\deffe, \effe\prangle |\in\ccc
\end{equation}
is continuous; moreover, it is bounded:
\begin{equation}
|\plangle\deffe, \uni (g)\cinque\effe\prangle-\plangle\deffe, \effe\prangle | \le \|\effe\|\tre  \|\deffe\| \bigg( 1 + \sup_{h\in G} \|\uni(h)\|\bigg).
\end{equation}
Therefore, since
\begin{eqnarray}
|\plangle\deffe, \unimt\cinque\effe\prangle-\plangle\deffe, \effe\prangle |
\spa & = & \spa \Big | \int_G \de\mut(g)\; \big( \plangle\deffe, \uni (g)\cinque\effe\prangle-\plangle\deffe, \effe\prangle
 \big)\Big | \nonumber\\
& \le & \spa \int_G \de\mut(g)\ |\plangle\deffe, \uni (g)\cinque\effe\prangle-\plangle\deffe, \effe\prangle |
\end{eqnarray}
and $\lim_{t\downarrow 0} \mut = \delta$ (weakly),
we conclude that
\begin{equation}
\lim_{t\downarrow 0} |\plangle\deffe, \unimt\cinque\effe\prangle-\plangle\deffe, \effe\prangle | = 0,\ \ \
\forall\quattro\effe\in\bsp,\ \forall\quattro\deffe\in\bspd.
\end{equation}
This completes the proof of the continuity of the
semigroup of operators $\{\unimt\}_{t\in\errep}$.~{$\square$}

\begin{remark} {\rm In the first part of the proof of Theorem~{\ref{semi-g}} we have implicitly used the fact that,
for any pair $\mu,\nu$ of Borel probability measures on $G$,
the formula
\begin{equation}
\int_G \de \mu\conv\nu (g)\;
f(g) =\int_G  \de \mu(g)\int_G\de \nu (h)\;f(gh),
\end{equation}
which holds
and for every continuous (real-valued) function $f$ with compact support on $G$, is satisfied by any bounded continuous
function too. In fact, by Urysohn's lemma~{\cite{Folland}} and the fact that $G$ is $\sigma$-compact,
there exists a sequence of compactly supported continuous functions
$\{\beta_n\}_{n\in\mathbb{N}}$ on $G$
such that $0\le \beta_n\le 1$ and  $\lim_{n\rightarrow\infty}\beta_n(g)=1$, for all $g\in G$. Therefore, for every bounded continuous
function $f$ on $G$, we have that
\begin{eqnarray}
\int_G  \de \mu(g)\int_G\de \nu (h)\;f(gh)
\spa & = & \spa \lim_{n\rightarrow\infty} \int_G  \de \mu(g)\int_G\de \nu (h)\; \beta_n(gh)\,f(gh)
\nonumber \\
& = & \spa \lim_{n\rightarrow\infty} \int_G \de \mu\conv\nu (g)\; \beta_n(g)\,f(g) = \int_G \de \mu\conv\nu (g)\;
f(g) ,
\end{eqnarray}
where we have used the `dominated convergence theorem'.~{$\blacksquare$}
}
\end{remark}

\begin{remark} {\rm
It is clear that $\{\unimtd\}_{t\in\errep}$ is the \emph{adjoint semigroup} of the semigroup of operators $\{\unimt\}_{t\in\errep}$.
Therefore, we have a remarkable case where the domain of continuity of the adjoint semigroup coincides with the whole
dual Banach space, property that does not hold in general, as first shown by Phillips~{\cite{Phillips}}.
Observe, moreover, that if $\uni$ is a representation,
then the semigroup $\{\unimtd\}_{t\in\errep}$ is generated by an antirepresentation
(i.e.\ $\uni^\ast$), and viceversa.~{$\blacksquare$}
}
\end{remark}

\begin{remark} {\rm
It can be shown by means of examples --- see Sect.~{\ref{examples}} --- that a randomly generated semigroup
may be generated by different pairs of the type $(\uni , \{\mut\}_{t\in\errep})$.~{$\blacksquare$}
}
\end{remark}

\section{Examples of randomly generated semigroups}
\label{examples}

In this section, we will consider three remarkable types of randomly generated semigroups, and we will
briefly comment the links connecting them.

\subsection{Random unitary semigroups}

In this case, we can identify the Banach space $\bsp$ and its dual $\bspd$ with a separable complex
Hilbert space $\hh$ --- note that we are now regarding the scalar product in $\hh$ as a pairing ---
and the uniformly bounded representation $\uni$ of a l.c.s.c.\ group $G$ with a unitary representation or
antirepresentation $U\colon G \rightarrow \mathcal{U}(\mathcal{H})$. Then, given a continuous convolution semigroup of
measures $\{\mut\}_{t\in\errep}$ on $G$, the randomly generated semigroup and the dual randomly generated
semigroup associated with the pair $(U,\{\mut\}_{t\in\errep})$ are given, respectively, by
\begin{equation}
\umt\cinque\phi=\intG U(g)\cinque\phi \; \de\mut(g),\ \ \ \umtd\phi = \intG U(g)^\ast \phi \; \de\mut(g),
\ \ \ \forall\quattro\phi\in\hh,
\end{equation}
where the integrals can be considered, in this case, as Bochner integrals (see the proof of Proposition~{\ref{separably}}),
and $U(g)^\ast$ is the `Hilbert space adjoint'
of $U(g)$. We will call these semigroups of operators \emph{random unitary semigroups}.
This terminology, which seems to be quite natural in this context, should however not confuse the reader for the fact that
it has been used in~{\cite{Aniello-brownian}} with a different meaning.

\subsection{Twirling semigroups}

We will now consider an example which is relevant for the theory
of open quantum systems~{\cite{Aniello-brownian}}.

Given a separable complex Hilbert space $\mathcal{H}$, we will
denote by $\opb$ a generic linear operator belonging to the Banach
space $\bH$ of bounded operators in $\hh$. The symbols $\opa$,
$\ops$ will denote generic operators in $\trc$ --- the Banach space
of trace class operators, endowed with the trace norm
--- and in the Hilbert-Schmidt space $\hs$
(which is a separable complex Hilbert space endowed with the Hilbert-Schmidt scalar product),
respectively. As is well known, we have that $\trc\subset\hs\subset\bH$ (strict inclusion,
for $\hh$ infinite-dimensional), and $\trc$, $\hs$ are $\ast$-ideals in $\bH$~{\cite{Reed}}.

We will identify the Banach space $\bsp$ of Sect.~{\ref{random}} with $\trc$.
The dual space of $\trc$ can
be identified with $\bH$ --- see~{\cite{Reed}} --- via the pairing
\begin{equation} \label{pairing}
\bH\times\trc\ni (\opb,\opa) \mapsto \tr\big(\opb\opa\big)\in\ccc.
\end{equation}
We will denote by $\supops$, $\dsupops$ the Banach algebras of bounded
(super)operators in $\trc$ and $\bH$, respectively (thus, to be identified
with $\bbsp$ and $\bbspd$).

Let $G$ be a l.c.s.c.\ group, and let $U$ be a projective
representation of $G$ in $\mathcal{H}$. The map
\begin{equation}
\rep \colon G\rightarrow\mathcal{U}(\hs)
\end{equation}
--- where $\mathcal{U}(\hs)$ is the unitary group of the Hilbert space $\hs$
--- defined by
\begin{equation} \label{definrep}
\rep(g)\hspace{0.3mm} \ops := U(g)\, \ops\, U(g)^\ast, \ \ \ \forall
\hspace{0.5mm}g\in G,\ \ \forall
\hspace{0.5mm}\ops\in\hs ,
\end{equation}
is a strongly continuous \emph{unitary} representation, even in the
case where the representation $U$ is genuinely \emph{projective};
see~{\cite{AnielloSP}}. Again, we stress that here $U(g)^\ast$ is the `Hilbert space adjoint'
of $U(g)$. Clearly, for every $g\in G$, the unitary
operator $\rep(g)$ in $\hs$ induces the Banach space isomorphism (a
surjective isometry) $\trc\ni\opa\mapsto\rep(g)\tre \opa\in\trc$.
Therefore, we can define the isometric representation
\begin{equation}
\urep\colon G \rightarrow\supops,\ \ \ \urep(g)\hspace{0.3mm} \opa
:= U(g)\, \opa\, U(g)^\ast, \ \ \ \forall \hspace{0.5mm}g\in G,\ \
\forall \hspace{0.5mm}\opa\in\trc.
\end{equation}
It turns out that the
the isometric representation $\urep$ of the l.c.s.c.\ group $G$ in
the Banach space $\trc$ is strongly continuous~{\cite{Aniello-brownian}}.

Now, given a continuous convolution semigroup $\{\mut\}_{t\in\errep}$ on $G$, we can define
the randomly generated semigroup associated with the pair $(\urep,\{\mut\}_{t\in\errep})$, i.e.,
the continuous semigroup of operators $\{\twis\}_{t\in\errep}\subset\supops$,
$\twis\equiv\urepmt$, with
\begin{equation} \label{concisa}
\twis \tre \opa = \int_G \de\mut(g)\ \urep(g)\tre\opa,\ \ \ \forall\quattro\opa\in\trc.
\end{equation}
Since the Banach space $\trc$ is separable, we can regard the integral in~{(\ref{concisa})} either
as a Pettis integral or as a Bochner integral. The corresponding dual randomly generated semigroup
$\{\dwis\}_{t\in\errep}\subset\dsupops$,
$\dwis\equiv\urepmtd$, is defined by
\begin{equation} \label{defidwis}
\dwis\tre \opb := \int_G \de\mut(g)\ U(g)^\ast\tre \opb\sei U(g) ,\ \ \
\forall\quattro\opb\in\bH,
\end{equation}
where the integral has to be regarded as a Gelfand integral.

The semigroup of operators $\{\twis\}_{t\in\errep}$ is a so-called \emph{twirling semigroup}~{\cite{Aniello-brownian}},
and it can be shown that it is a quantum dynamical semigroup; namely, a continuous semigroup of operators consisting of
positive, trace-preserving, bounded linear maps in $\trc$, whose adjoints
(acting in the Banach space $\bH$) are completely
positive~{\cite{Holevo}}.
Thus, it describes the evolution of an open quantum system
in the Schr\"odinger picture, while the dual semigroup $\{\dwis\}_{t\in\errep}$ describes the corresponding evolution in the
Hiesenberg picture~{\cite{Holevo,Breuer}}. We remark that a twirling semigroup
may be generated by different pairs of the type $(\urep , \{\mut\}_{t\in\errep})$. In fact,
if $\hh$ is finite-dimensional, one can always assume that
$U$ is the defining representation of a special unitary group $\sunig(n)$,
$n=\dim(\hh)$, and $ \{\mut\}_{t\in\errep}$
a (generic) continuous convolution semigroup on $\sunig(n)$,  see~{\cite{Aniello-brownian}}.

Finally, it is interesting to note that one can define
the randomly generated semigroup $\{\twishs\}_{t\in\errep}$ associated with the pair $(\rep,\{\mut\}_{t\in\errep})$
--- i.e., $\twishs\equiv\repmt$ ---
which acts in the Hilbert space $\hs$:
\begin{equation} \label{defitwihs}
\twishs\tre \ops := \int_G \de\mut(g)\ \rep(g)\tre\ops ,\ \ \
\forall\quattro\ops\in\hs.
\end{equation}
Here, again, we can regard the integral in~{(\ref{defitwihs})} either
as a Pettis integral or as a Bochner integral. We stress that
the definition of the corresponding dual semigroup is independent of the choice
of one of the two natural pairings $(\ops,\opsp) \mapsto \tr\big(\ops\quattro\opsp\big)$, or
$(\ops,\opsp) \mapsto \tr\big(\ops^\ast\opsp\big)$ (Hilbert-Schmidt scalar product), for identifying $\hs$
with its dual space. Also note that the semigroup of operators
$\{\twis\}_{t\in\errep}$ can be considered as the restriction to $\trc$ of the semigroup
$\{\twishs\}_{t\in\errep}$.

\subsection{Tomographic semigroups}
\label{tomographic}

Let $G$ be a l.c.s.c.\ group and $\mG$
the left Haar measure on $G$ (which is, of course, unique up to normalization),
and let us set $\ldg\equiv\lgc$. We will denote by $\modu$ the modular function on $G$.
Given a multiplier $\emme$ for $G$, consider the map
$\twoside\colon G\rightarrow \mathcal{U}(\ldg)$
defined by
\begin{equation} \label{two-sided}
\big(\twoside(g)\tre f\big)(h) := \modu (g)^{\frac{1}{2}}\hspace{0.9mm}
\temme(g,h)\hspace{0.8mm} f(g^{-1}h g) ,\ \ \ f\in\ldg,
\end{equation}
where the function $\temme\colon G\times G\rightarrow \mathbb{T}$ is
defined as follows:
\begin{equation}
\temme(g,h) := \emme (g,g^{-1} h)^\ast \tre \emme (g^{-1} h,g), \ \ \ \forall\quattro g,h\in G.
\end{equation}
The map $\twoside$ is a strongly continuous unitary representation,
see~{\cite{AnielloSP}}. Then, given a continuous convolution semigroup $\{\mut\}_{t\in\errep}$ on $G$, we can define
the randomly generated semigroup associated with the pair $(\twoside,\{\mut\}_{t\in\errep})$, namely, the
random unitary semigroup $\{\smg\equiv\twomt\}_{t\in\errep}\subset\mathcal{B}(\ldg)$ determined by
\begin{equation}
\smg f =\intG  \twoside(g)\tre f \; \de\mut(g),
\ \ \ \forall\quattro  f\in\ldg,
\end{equation}
where the integral can be considered as a  Bochner integral.

Is there any link connecting the random unitary semigroup $\{\smg\}_{t\in\errep}$ with
the previously defined twirling
semigroups $\{\twis\}_{t\in\errep}$ and $\{\twishs\}_{t\in\errep}$?

The answer is positive if we assume that the projective representation
$U\colon G\rightarrow\mathcal{U}(\mathcal{H})$, with multiplier $\emme$,
that allows us to define the twirling semigroups is a \emph{square integrable} irreducible representation~{\cite{Aniello}}.
In this case, one can define an isometric linear operator
\begin{equation}
\mathcal{W} \colon \hs\rightarrow\ldg,
\end{equation}
the so-called
\emph{tomographic map}, or (generalized) \emph{Wigner map},
generated by $U$~{\cite{AnielloSP}}. \\
At this point, we will state --- without proofs and in a rather sketchy way --- a few facts.
A detailed exposition is beyond the aims of the present contribution and will be given elsewhere~{\cite{Aniello-preparation}}.
It turns out that the tomographic map
$\wigu$ intertwines the extended twirling semigroup
$\{\twishs\}_{t\in\errep}$ with the semigroup
$\{\smg\}_{t\in\errep}$, i.e.,
\begin{equation} \label{intertpro}
\wigu\sei \twishs  = \smg\tre\wigu,\ \ \ \forall \quattro t\in\errep .
\end{equation}
Therefore, the range $\ran (\wigu)$ of the tomographic map
$\wigu$ is stable under the action of the semigroup of operators
$\{\smg\}_{t\in\errep}$, namely,
\begin{equation}
\smg\cinque \ran (\wigu)\subset\ran (\wigu),\ \ \ \forall \quattro
t\in\errep .
\end{equation}
Similarly, the linear subspace $\wigu(\trc)$ of $\ran (\wigu)$ is stable
under the action of $\{\smg\}_{t\in\errep}$, and the restriction of the tomographic map
to $\trc\subset\hs$ intertwines the twirling semigroup $\{\twis\}_{t\in\errep}$ with the restriction
of the semigroup of operators $\{\smg\}_{t\in\errep}$ to $\wigu(\trc)$.

We may call the randomly generated semigroup $\{\smg\}_{t\in\errep}$ ---
due to the context where it arises in a natural way --- a
\emph{tomographic semigroup} in $\ldg$, associated with the
multiplier $\emme$. We stress, however, that this semigroup of operators is well
defined independently of the existence of a square integrable
representation of $G$ with multiplier $\emme$ (hence, of a
tomographic map).

\section{Conclusions}
\label{conclusions}

In the present contribution, we have considered an interesting class of semigroups
of operators: the \emph{randomly generated semigroups}.
The examples discussed in Sect.~{\ref{examples}} show that this class encompasses,
in a unique mathematical framework, various types of
semigroups of operators that may look, at a first glance, quite different.
Observe indeed that --- using the notation of Subsect.~{\ref{tomographic}} --- if the l.c.s.c.\ group $G$
admits a square integrable representation $U$, with multiplier $\emme$, then the restriction
of the tomographic semigroup $\{\smg\}_{t\in\errep}$ to the range of the tomographic map $\wigu$ (generated by $U$)
is \emph{mutatis mutandis} --- with a space of $\ccc$-valued functions (`tomograms')
isomorphically replacing a space of operators
--- a twirling semigroup. A simple, but very interesting, example
is the case where $G=\erre^n$ (the $n$-dimensional Lie group of translations)
and $U$ is a \emph{Weyl system}~{\cite{Weyl}}, namely, a suitable infinite-dimensional,
irreducible projective representation of
$\erre^n$. As the reader may easily guess,
this case is related with the dynamics of an open quantum system `in the phase
space formulation' {\it \`a la} Weyl-Wigner.
This example will be discussed in a forthcoming paper~{\cite{Aniello-preparation}},
where the notion of tomographic semigroup will be studied in detail.

Finally, we note that a further example of a type of randomly generated semigroups
is provided by the \emph{probability semigroups}~{\cite{Aniello-brownian,Heyer}}. Let $G$ be a l.c.s.c.\
group and $\{\mut\}_{t\in\errep}$ a continuous convolution semigroup of measures on $G$. Then, one can define
a continuous semigroup of operators $\{\prop_t\}_{t\in\errep}$ in $\cz(G)$ ---
the Banach space of all continuous
$\erre$-valued functions on $G$ vanishing at infinity, endowed with the `sup-norm' ---
by setting
\begin{equation}
\big(\prop_t\tre f\big)(g) := \int_G f(gh)\; \de\mut(h) , \ \ \ \forall\quattro f\in \cz(G).
\end{equation}
The verification that $\{\prop_t\}_{t\in\errep}$ is a randomly generated semigroup and the determination
of the associated adjoint semigroup is an interesting exercise that we leave to the reader.

\section*{Acknowledgments}

The main results of the paper were presented by the author at the
international conference {\it The 42nd Symposium on Mathematical Physics,
``Quantum Information, Quantum Channels -- Theory and Applications''}
(19-22 June 2010, Toru\'n, Poland).
The author wishes to thank the organizers for their very kind hospitality.



\end{document}